\documentclass[aps,prl,reprint,superscriptaddress,longbibliography,nofootinbib]{revtex4-1}
\usepackage{lipsum}

\usepackage[utf8]{inputenc}

\usepackage{amsfonts}
\usepackage{amsmath}
\usepackage{tikz}
\usetikzlibrary{arrows,calc}
\usetikzlibrary{patterns,snakes}

\usepackage{ upgreek }

\def\g{{\rm g}}
\def\C{{\mathcal C}}

\newcommand{\R}{\ensuremath{\mathbb R}}

\newcommand{\Z}{\ensuremath{\mathbb Z}}

\DeclareMathOperator{\Tr}{Tr}

\newcommand{\bea}{\begin{eqnarray}}
\newcommand{\eea}{\end{eqnarray}}

\begin{document}

%\preprint{}

\title{Solutions of modular bootstrap constraints from quantum codes}
%% Conformal Field Theories from Quantum Codes 
%% Isospectral and Fake Conformal Field Theories from Quantum Codes: Challenges for the Modular Bootstrap Program


\author{Anatoly Dymarsky}
\affiliation{Department of Physics and Astronomy, \\ University of Kentucky, Lexington, KY 40506\\[2pt]}
\affiliation{Skolkovo Institute of Science and Technology, \\ Skolkovo Innovation Center, Moscow, Russia, 143026\\[2pt]}
\author{Alfred Shapere}
\affiliation{Department of Physics and Astronomy, \\ University of Kentucky, Lexington, KY 40506\\[2pt]}


\date{\today}
\begin{abstract}
Modular invariance imposes rigid constraints on the partition functions of two-dimensional conformal field theories. Many fundamental results follow strictly from modular invariance and unitarity, giving rise  to the numerical  modular  bootstrap program.  Here we report on a way to relate a particular family of quantum error correcting codes to a family of  ``code CFTs'', which forms a  subset of the space of Narain CFTs. This correspondence reduces modular invariance of the 2d CFT partition function to a few simple algebraic relations obeyed by a multivariate polynomial characterizing  the corresponding code. Using this correspondence we construct many explicit examples of physically distinct isospectral theories, as well as many examples of nonholomorphic functions, which satisfy all the basic properties of a 2d CFT partition function yet are not associated with any known CFT.

\end{abstract}

%\keywords{}

\maketitle

Two-dimensional conformal field theories (CFTs) enjoy an exceptionally wide range of applications from condensed matter physics to string theory and quantum gravity. Characterizing the space of all CFTs is one of the central tasks of the conformal modular bootstrap program \cite{Hellerman,hellerman2011bounds,keller2013modular,friedan2013constraints,qualls2014bounds,
hartman2014universal,qualls2015universal,kim2016reflections,lin20172,anous2018parity,collier2018modular,afkhami2019fast,
cho2019genus,hartman2019sphere,afkhami2020high,afkhami2020free,gliozzi2020modular}, which aims to deduce universal properties of 2d theories, as well as details of specific models, from the modular invariance and non-negativity of their partition functions on the torus. In a nutshell, the modular bootstrap conditions form a proper subset of the conditions for conformal invariance and unitarity, and hence a solution to the bootstrap equations  does not necessarily imply the existence of an actual theory.  Nevertheless it has been observed numerous times, including  in the context of  the conformal bootstrap in $d>2$, that a robust solution of the bootstrap constraints, {\it e.g.}~a ``kink'' in the exclusion plot, reflects the presence of an actual  theory. This picture is consistent with another observation, that with the exception of a limited family of examples related to chiral models \cite{schellekens1992meromorphic} and a class of candidate partition functions for rational CFTs with two characters \cite{mukhi2019classification} all currently known non-chiral candidate partition  functions -- nonholomorphic modular-invariant functions $Z(\tau,\bar\tau)$ which can be expanded in (Virasoro) characters with non-negative integral coefficients and with leading coefficient equal to one (reflecting the requirement of a unique CFT vacuum) -- are partition functions of actual 2d theories. Furthermore, in practice, solutions of the modular bootstrap equations are typically associated with specific theories, which implicitly assumes that distinct CFTs must have different partition functions.

We show  that this simple picture is not accurate, and the true situation is maximally complex: (i) there are many functions $Z(\tau,\bar\tau)$ which are not partition functions of any (known) theory, and (ii)  there are many examples of isospectral physically distinct theories, {\it i.e.}~pairs, triplets, and even groups of 11 theories, which are all different but share the same partition function.  Our results therefore expose the limitations of any approach, including the modular bootstrap program, which aims to characterize CFTs solely on the basis of their torus partition functions.  

Our construction is very explicit and uses a map from a particular class of quantum error correcting codes to the space of 2d Narain CFTs that describe the compactification of free scalar fields on a multi-dimensional cube. The CFT partition function in this case is fully determined by the code, namely by the code's refined enumerator polynomial, introduced below. A direct search in the space of $n$-qubit codes with $n\leq 8$ readily reveals many dozens of distinct codes with the same enumerator, leading to many examples  of physically distinct isospectral theories. There are also many polynomials obeying the standard symmetries and constraints of the enumerator polynomial, which are nevertheless not enumerator polynomials of any actual code. These ``fake'' polynomials provide thousands of examples of modular invariant $Z(\tau,\bar \tau)$ which are sums of $U(1)^n\times U(1)^n$ characters,
\bea
\label{expansion}
Z(\tau,\bar \tau)={1+\sum_{h,\bar h} C_{h,\bar h}\, q^h {\bar q}^{\bar h}\over |\eta(\tau)|^{2n}},\qquad q=e^{2\pi i \tau},
\eea 
with non-negative integral coefficients $C_{h,\bar h}$, which are not partition functions of any known CFTs.

We start by reviewing graph codes, a particular family of real self-dual stabilizer codes \cite{schlingemann2001quantum,schlingemann2002stabilizer}.
A graph code is specified by a binary symmetric matrix $\rm B$ -- the adjacency matrix of an unoriented graph on $n$ nodes, ${\rm B}_{ii}=0$, ${\rm B}_{ij}={\rm B}_{ji}\in \{0,1\}$.  The adjacency matrix defines a set of $n$ generators  
\bea
\label{graph}
\g_i=\sigma_x^i \prod_{j=1}^n \left(\sigma_z^j\right)^{{\rm B}_{ij}} \label{g}
\eea
acting on the space of $n$ qubits.
Because the matrix $\rm B$ is symmetric, the generators $\g_i$ commute. Furthermore they are nilpotent $\g_i^2={\rm I}$ and they generate an Abelian stabilizer group $\cal S$ of rank $2^n$. There is a unique (up to scalar rescaling) state $\psi_\C$ which is invariant under the action of any  element of the stabilizer group, $\g_i \psi_\C=\psi_\C$.  Known as the graph state \cite{dur2003multiparticle,glynn2004geometry,van2004graphical,hein2006entanglement}, $\psi_\C$ can be written explicitly in terms of the ``computational'' up-down basis.
Self-duality of the code implies that the Abelian stabilizer group possesses the maximal possible number of independent generators. The code is called real because all matrix elements of \eqref{g} are real. 

The most general element of the stabilizer group is a product of generators 
\bea
\g(\alpha)=\prod_{i=1}^n \g_i^{\alpha_i},
\eea
characterized by a binary vector $\vec{\alpha}\in \Z_2^n$. Up to a sign it can be written as a product of Pauli operators 
\bea
\g(\alpha)=\epsilon \prod_{i=1}^n (\sigma_x^i)^{\alpha_i}  \prod_{j=1}^n (\sigma_z^j)^{\beta_j}, \quad \epsilon=\pm 1, \label{ga}
\eea
where the binary vector $\vec{\beta}:={\rm B}\,\vec{\alpha}\, \, {\rm mod}\, \,  2$. With this definition $w_y(\alpha)=\vec{\alpha}\cdot \vec{\beta}$ counts the number of  Pauli matrices $\sigma_y$ in \eqref{ga}, while $w(\alpha)=\sum_i \alpha_i+\beta_i-w_y(\alpha)$ counts the total number of qubits on which $\g(\alpha)$ acts non-trivially. 
A basic characteristic of a code is its enumerator polynomial, which counts the number of $\g \in {\cal S}$ which act on a particular number of qubits. For our purposes we consider the closely related refined enumerator polynomial, which keeps track of the total number of affected qubits, as well as the number of $\sigma_y$'s,
\bea
W(x.y,z)=\sum_{\vec{\alpha}\in \Z_2^n} x^{n-w(\alpha)} y^{w_y(\alpha)} z^{w(\alpha)-w_y(\alpha)}. \label{REP}
\eea
When the code is real, there is always an even number of $\sigma_y$ in each $\g(\alpha)$, and therefore $W$ is invariant under
\bea
\label{y}
y \rightarrow -y. \label{y}
\eea
Furthermore,  the refined enumerator polynomial of a self-dual code is 
 invariant under the transformation 
\bea
x\to \frac{x+y+2 z}{2},\quad y\to \frac{x+y-2 z}{2},\quad z\to \frac{x-y}{2}.\quad\ 
\label{duality}
\eea
This symmetry follows from the MacWilliams identity \cite{macwilliams1978self,shor1997quantum,nebe2006self}.

In the context of quantum codes it is natural to call codes equivalent if they are related by a permutation of qubits, which at the level of graphs is simply a relabeling of nodes. Two codes are also said to be equivalent if they are related by a local Clifford (LC) transformation, a unitary transformation $\g_i\rightarrow U\, \g_i\, U^\dagger$ acting on the individual qubits $U=u_1\otimes \dots \otimes u_n$, which preserves the form of the stabilizer generators as tensor products of Pauli operators.  If we restrict attention to real codes, the only allowed LC transformations are those generated by the Hadamard matrix, $u_i=H$, which simply exchanges $\sigma_x^i \leftrightarrow \sigma_z^i$. 
Following an LC transformation, the generators can be recombined to bring them again to graph form \eqref{graph}, so that the action of the LC group can be understood in terms of graph transformations. At the level of graphs, all code equivalence transformations among real codes generate an orbit in the space of graphs under the action of edge local complementation (ELC) \cite{van2005edge}. 
The action of an ELC transformation on the graph adjacency matrix is as follows 
\bea
{\rm B} \rightarrow \left((D+{\rm I}){\rm B}+D\right)\left(D\, {\rm B}+D+{\rm I}\right)^{-1},\label{elcp}
\eea 
where all operations, including matrix inversion, are understood mod 2, and $D$ is an arbitrary diagonal matrix. 

Clearly, permutations of qubits and exchanges $\sigma_x^i \leftrightarrow \sigma_z^i$ do not change \eqref{REP}; therefore two graph codes associated with graphs related by ELC will have the same 
refined enumerator polynomial.

In the context of quantum computation, real codes are not special and code equivalence is usually defined to include the full group of LC transformations. At the level of graphs, the full equivalence group gives rise to orbits under local complementation. A classification of orbits under local complementation for all graphs on $n\leq 12$ nodes has been performed in \cite{danielsen2006classification}, where it was used to enumerate all equivalence classes of self-dual stabilizer codes for $n\leq 12$ qubits. The  orbits of graphs under ELC are suborbits within the orbits of local complementation. To our knowledge they have not been fully classified previously. We classify all ELC orbits for graphs on  $n\leq 8$ nodes.

At this point we would like to assign to each stabilizer group of the form \eqref{graph} a Narain CFT, which describes $n$ free scalar fields compactified on an $n$-dimensional cube of ``unit'' size $2\pi$ in the presence of quantized $B$-field flux. The Narain CFT can be defined in terms of an even self-dual Lorentzian lattice in $\R^{n,n}$. Starting from \eqref{graph}, or equivalently from the graph adjacency matrix $\rm B$, we define a lattice  generator matrix 
\bea
\Uplambda=\left(\begin{array}{c|c}
2\, {\rm I}\,  &\, \, B\, \,  \\ \hline
0\, &\,  {\rm I}\end{array}\right)/\sqrt{2}, \label{Lambda}
\eea
where $B_{ij}$ is an arbitrary antisymmetric matrix satisfying 
\bea
{\rm  B}=B\, \, {\rm mod}\, \, 2. \label{B}
\eea
It is easy to check that \eqref{Lambda} satisfies 
$\Uplambda^T g\,  \Uplambda=g,$
where the Lorentzian metric is
\bea
g=\left(\begin{array}{c|c}
\, 0\,  &\, {\rm I}\, \\ \hline
{\rm I} & 0 
\end{array}\right). \label{Mink}
\eea
Different values of $B$ satisfying \eqref{B} lead to the same Lorentzian lattice, and the corresponding CFTs are related by T-duality. The relation between graph codes and Narain CFTs can be extended to all real self-dual codes \cite{QLC}. We will call the corresponding theories code CFTs. 

The partition function of a code CFT is fully specified by the underlying code itself, or more precisely by its refined enumerator polynomial
\bea
Z(\tau,\bar\tau)={W_\C\left({b\,{\bar b}+c\,{\bar c}},{b\,{\bar b}-c\, {\bar c}},{a\, {\bar a}}\right)\over 2^n |\eta(\tau)|^{2n}}, \label{Z}
\eea
where $a=\theta_2(\tau),\quad b=\theta_3(\tau),\quad c=\theta_4(\tau).$
Using standard relations for the Jacobi theta functions one can verify that \eqref{Z} is invariant under $\tau \rightarrow \tau+1$ due to \eqref{y} and under $\tau \rightarrow -1/\tau$ due to \eqref{duality}. This ensures modular invariance of  the code CFT partition function. The refined enumerator polynomial satisfies other  conditions, which follow from its definition: $W$ should be homogeneous, $W(1,0,0)=1$,  and all coefficients of the polynomial $W(x,y,z)$ should be non-negative integers. At the level
of the partition function this ensures  $Z(\tau,\bar \tau)$ is of the form \eqref{expansion} with non-negative integral $C_{h,\bar h}$.

The condition of invariance of the refined enumerator polynomial under \eqref{y} and \eqref{duality} is easy to ``solve'' in full generality. A polynomial $W(x,y,z)$ invariant under both symmetries is an arbitrary polynomial in three generating polynomials,
\bea
\nonumber
W_1=x+z,\quad W_2=x^2+y^2+2z^2,\\ W_3=x^3+3 x y^2+4 z^3.\qquad
\label{invpols}
\eea
Imposing other conditions, $W(1,0,0)=1$ and non-negativity of integer coefficients, reduces the problem of finding such a $W$ to an exercise in discrete linear programming, which can be easily solved using computer algebra for small and moderate $n$. As is well known in the context of classical codes \cite{nebe2006self}, there are many more polynomials satisfying the aforementioned conditions than there are actual codes. We call such polynomials that do not correspond to codes ``fake'' refined enumerator polynomials. Their number grows  quickly with $n$. There are no fake polynomials for $n=1$ and $n=2$, there are 6 fake polynomials for $n=3$,

\bea
W&=&x^3+2 x^2 z+3 x z^2+y^2 z+z^3,\\
W&=&x^3+x^2 z+3 x z^2+2 y^2 z+z^3,\\
W&=&x^3+2 x^2 z+x y^2+2 x z^2+2 z^3,\\
W&=&x^3+x y^2+2 x z^2+2 y^2 z+2 z^3,\\
W&=&x^3+x^2 z+2 x y^2+x z^2+3 z^3,\\
W&=&x^3+2 x y^2+x z^2+y^2 z+3 z^3,
\eea
there are 11 for $n=4$, 128 for $n=5$, $2835$ for $n=6$, $71164$ for $n=7$, 4012529 for $n = 8$, and so on. Each fake enumerator polynomial defines a candidate partition function via \eqref{Z} which is modular invariant and non-negative, yet which is not the partition function of any known physical theory. The multitude of ``fake'' partition functions of small central charge that do not correspond to known CFTs is one of our main results. 

\begin{figure}[t]
\centering
\includegraphics[width=0.5\textwidth]{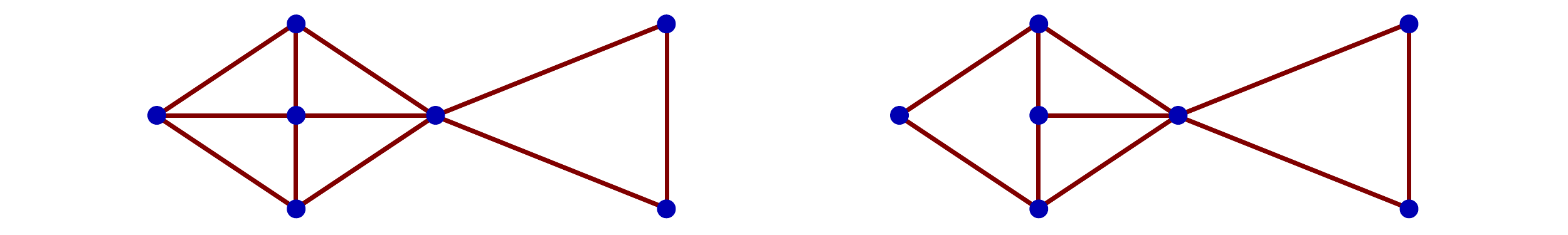}
\caption{``Fish'' graphs -- representatives of two ELC orbits of graphs, which at the level of graph codes share the same refined enumerator polynomial \eqref{rep7} and lead to a pair of isospectral non-chiral CFTs with $c={\bar c}=7$.}
\label{fig:fish}
\end{figure} 

T-duality is a symmetry of any Narain CFT; it leaves the physical theory intact, but can change the lattice generator matrix.  In principle T-duality can map one code theory into another code theory. It is easy to see that the code equivalence transformations that respect the reality condition -- permutations of qubits and exchanges of $\sigma_x^i \leftrightarrow \sigma_z^i$ -- become T-duality transformations, at the level of lattices and their associated Narain theories. The opposite is also true: any T-duality transformation which maps a code theory into another code theory is, at the level of codes, a code equivalence which involves qubit permutations and exchanges $\sigma_x^i \leftrightarrow \sigma_z^i$ \cite{QLC}. We therefore arrive at a key observation: two code CFTs associated with graph codes are T-dual to each other (physically equivalent) if and only if the corresponding graphs are related by edge local complementation. Our classification of all ELC orbits for graphs on $n\leq 8$ nodes therefore provides an enumeration of all physically distinct code CFTs. By comparing refined enumerator polynomials associated with different ELC orbits we find no degeneracies for $n\leq 6$ and one degeneracy for $n=7$. Namely, one refined enumerator 
\bea
\label{rep7}
&& W=x^7 + x^5 y^2 + 5 x^4 y^2 z + 5 x^2 y^4 z + x^5 z^2+ \\
\nonumber 
&& \ \ 12 x^3 y^2 z^2 + 9 x y^4 z^2 + 4 x^4 z^3 + 22 x^2 y^2 z^3 + 4 y^4 z^3+\\  
\nonumber
&& \ \ 5 x^3 z^4 + 25 x y^2 z^4 + 11 x^2 z^5  + 11 y^2 z^5 + 10 x z^6 + 2 z^7. 
\eea
is the same for two different classes of codes, giving rise to a pair of physically distinct isospectral CFTs. 
At the level of lattices, this corresponds to a pair of isospectral even self-dual Lorentzian lattices (also equipped with a Euclidean metric), analogous to Milnor's example of isospectral even self-dual lattices in $\R^{16}$ \cite{conway1997sensual}.
At the level of graphs, the two ELC classes respectively include 9 and 10 different graphs (up to isomorphisms). We have chosen one graph from each class based on simplicity and aesthetics, and have depicted them in Fig.~\ref{fig:fish}. 

For $n=8$, there are already $60$ new pairs and $5$ triplets of isospectral code theories, too many to list here. While we did not fully analyze ELC classes for higher $n$, we can utilize the analysis of \cite{danielsen2006classification}    
classifying orbits under local complementation. Since any ELC orbit is a subset within an orbit under local complementation, graphs belonging to different local complementation orbits correspond to physically different theories. Going through the values $n=9-11$ we find many examples of pairs, triples, and in fact $k$-tuples of isospectral code theories for every $k\leq 11$, confirming the expectation that the number of isospectral theories grows rapidly with $n$. These findings raise the question of identifying a mechanism that could explain the growing number of isospectral theories.

A natural generalization of our construction would be to consider the grand canonical partition function decorated by $U(1)^n \times U(1)^n$ charges, $Z(\tau,\bar \tau,\xi,\bar \xi)$, where $\xi_I$ and $\bar \xi_I$ are the associated chemical potentials. The choice $\xi_I=\xi$ and $\bar \xi_I=\bar \xi$ for all $I=1,\dots,n$ preserves permutation symmetry. With this choice, for code theories the grand canonical partition function is given in terms of the {\it full} code  enumerator polynomial $W(t,x,y,z)$ by an expression generalizing \eqref{Z}
\bea
Z(\tau,\bar\tau,\xi,\bar \xi)={W\left(b \bar b+c\bar c, a\bar a+d\bar d, b \bar b-c\bar c, a\bar a-d\bar d\right)\over 2^n |\eta(\tau)|^{2n}}\qquad  \label{refinedZ}
\eea
where $a=\theta_2(\tau,\xi),\ b=\theta_3(\tau,\xi),\ c=\theta_4(\tau,\xi),\ d=\theta_1(\tau,\xi)$.
Just as one finds for $Z(\tau,\bar\tau)$ in \eqref{Z},
%As in the conventional case, 
modular covariance of $Z(\tau,\bar\tau,\xi,\bar \xi)$ follows from the MacWilliams identity for $W(t,x,y,z)$. The introduction of nonzero $\xi,\bar\xi$ removes the degeneracy of $Z$ between the isospectral but not T-dual $n=7$ theories associated with the graphs shown in Fig.~\ref{fig:fish}. A detailed derivation of \eqref{refinedZ} together with a discussion of the $n=7$ case can be found in the Supplemental Material at the [url], which includes Ref.~\cite{kraus2007partition}.

The construction relating quantum codes to rational CFTs outlined in this paper is not unique. Other classes of codes can be naturally mapped to families of non-chiral rational theories \cite{future}. As was shown in \cite{witten1989quantum}, conformal blocks of rational CFTs are wave-functions of the dual Chern-Simons theory, which form a finite-dimensional Hilbert space. Provisionally we identify this Hilbert space with the Hilbert space of qubits the quantum code describes.  Exploring further the role of quantum codes in CFTs and in the holographic correspondence is an important program to which we hope to return in the future. 

There is a simpler {\it chiral} version of our  construction. Classical binary doubly-even self-dual codes are associated with even self-dual lattices, which can be used to define chiral CFTs. As is well known, starting with $n=24$ there exist ``fake'' enumerator polynomials, and hence would-be  holomorphic  partition functions $Z(\tau)$ which are known to have no CFT counterpart \cite{schellekens1992meromorphic}. Furthermore,  starting in dimension $n=16$, there are isospectral self-dual lattices and hence isospectral physically inequivalent chiral theories \cite{PhysRevD.35.648,NARAIN198641,NARAIN1987369}. The main difference of our work is that it applies to non-chiral theories, which are much less studied. Furthermore, our examples arise in large numbers at much smaller values of the central charge, with implications for the numerical conformal bootstrap. 

As a final remark we notice that the partition function \eqref{Z} is a polynomial in code CFT partition functions associated with the polynomials \eqref{invpols} \cite{QLC}. This observation suggests a simple way to construct many new modular-invariant $Z(\tau,\bar \tau)$ by simply taking polynomials of any collection of CFT partition functions and making sure that the coefficients in the character expansion are non-negative integers and the vacuum character is unique. For homogeneous polynomials, each term has leading small-$q$ behavior $q^{-c/12}$, so uniqueness of the vacuum is a constraint on the coefficients of the polynomial. The examples in this paper show that this constraint has many nontrivial solutions. For an inhomogeneous polynomial in a single variable, uniqueness of the vacuum requires the coefficient of the leading term to be one, as in   \cite{witten2007three}. We expect that there are many more possibilities for inhomogenious polynomials in several variables, leading to a plethora of modular-invariant $Z(\tau,\bar \tau)$ not associated with any CFT.

{\it Conclusions.} In this paper we have reported on a way to associate graph codes, a family of quantum error correcting codes, to a specific class of Narain CFTs, whose elements we call code theories. Code CFTs describe free scalar fields compactified on a multidimensional cube in the presence of quantized B-flux. This mapping between graph codes and CFTs provides a new way to study non-chiral theories. It allows us to construct many explicit examples of isospectral non-chiral theories, as well as many examples of would-be partition functions \eqref{expansion} which do not correspond to any known theories. 
These examples emphasize the fact that solutions of the modular bootstrap equations need not correspond to a unique CFT, nor indeed to any CFT at all. 

Many  technical details relevant to our presentation,  including data necessary to construct pairs and triples of $n=8$ isospectral theories, can be  found in the accompanying work \cite{QLC}.

\acknowledgments
We thank Petr Kravchuk and Xi Yin for discussions. AD is supported by the National Science Foundation under Grant No.~PHY-2013812.

\bibliography{QLC}

\begin{thebibliography}{39}%
\makeatletter
\providecommand \@ifxundefined [1]{%
 \@ifx{#1\undefined}
}%
\providecommand \@ifnum [1]{%
 \ifnum #1\expandafter \@firstoftwo
 \else \expandafter \@secondoftwo
 \fi
}%
\providecommand \@ifx [1]{%
 \ifx #1\expandafter \@firstoftwo
 \else \expandafter \@secondoftwo
 \fi
}%
\providecommand \natexlab [1]{#1}%
\providecommand \enquote  [1]{``#1''}%
\providecommand \bibnamefont  [1]{#1}%
\providecommand \bibfnamefont [1]{#1}%
\providecommand \citenamefont [1]{#1}%
\providecommand \href@noop [0]{\@secondoftwo}%
\providecommand \href [0]{\begingroup \@sanitize@url \@href}%
\providecommand \@href[1]{\@@startlink{#1}\@@href}%
\providecommand \@@href[1]{\endgroup#1\@@endlink}%
\providecommand \@sanitize@url [0]{\catcode `\\12\catcode `\$12\catcode
  `\&12\catcode `\#12\catcode `\^12\catcode `\_12\catcode `\%12\relax}%
\providecommand \@@startlink[1]{}%
\providecommand \@@endlink[0]{}%
\providecommand \url  [0]{\begingroup\@sanitize@url \@url }%
\providecommand \@url [1]{\endgroup\@href {#1}{\urlprefix }}%
\providecommand \urlprefix  [0]{URL }%
\providecommand \Eprint [0]{\href }%
\providecommand \doibase [0]{http://dx.doi.org/}%
\providecommand \selectlanguage [0]{\@gobble}%
\providecommand \bibinfo  [0]{\@secondoftwo}%
\providecommand \bibfield  [0]{\@secondoftwo}%
\providecommand \translation [1]{[#1]}%
\providecommand \BibitemOpen [0]{}%
\providecommand \bibitemStop [0]{}%
\providecommand \bibitemNoStop [0]{.\EOS\space}%
\providecommand \EOS [0]{\spacefactor3000\relax}%
\providecommand \BibitemShut  [1]{\csname bibitem#1\endcsname}%
\let\auto@bib@innerbib\@empty
%</preamble>
\bibitem [{\citenamefont {Hellerman}(2011)}]{Hellerman}%
  \BibitemOpen
  \bibfield  {author} {\bibinfo {author} {\bibfnamefont {Simeon}\ \bibnamefont
  {Hellerman}},\ }\bibfield  {title} {\enquote {\bibinfo {title} {A universal
  inequality for cft and quantum gravity},}\ }\href {\doibase
  10.1007/JHEP08(2011)130} {\bibfield  {journal} {\bibinfo  {journal} {Journal
  of High Energy Physics}\ }\textbf {\bibinfo {volume} {2011}},\ \bibinfo
  {pages} {130} (\bibinfo {year} {2011})}\BibitemShut {NoStop}%
\bibitem [{\citenamefont {Hellerman}\ and\ \citenamefont
  {Schmidt-Colinet}(2011)}]{hellerman2011bounds}%
  \BibitemOpen
  \bibfield  {author} {\bibinfo {author} {\bibfnamefont {Simeon}\ \bibnamefont
  {Hellerman}}\ and\ \bibinfo {author} {\bibfnamefont {Cornelius}\ \bibnamefont
  {Schmidt-Colinet}},\ }\bibfield  {title} {\enquote {\bibinfo {title} {Bounds
  for state degeneracies in 2d conformal field theory},}\ }\href@noop {}
  {\bibfield  {journal} {\bibinfo  {journal} {Journal of High Energy Physics}\
  }\textbf {\bibinfo {volume} {2011}},\ \bibinfo {pages} {127} (\bibinfo {year}
  {2011})}\BibitemShut {NoStop}%
\bibitem [{\citenamefont {Keller}\ and\ \citenamefont
  {Ooguri}(2013)}]{keller2013modular}%
  \BibitemOpen
  \bibfield  {author} {\bibinfo {author} {\bibfnamefont {Christoph~A}\
  \bibnamefont {Keller}}\ and\ \bibinfo {author} {\bibfnamefont {Hirosi}\
  \bibnamefont {Ooguri}},\ }\bibfield  {title} {\enquote {\bibinfo {title}
  {Modular constraints on calabi-yau compactifications},}\ }\href@noop {}
  {\bibfield  {journal} {\bibinfo  {journal} {Communications in Mathematical
  Physics}\ }\textbf {\bibinfo {volume} {324}},\ \bibinfo {pages} {107--127}
  (\bibinfo {year} {2013})}\BibitemShut {NoStop}%
\bibitem [{\citenamefont {Friedan}\ and\ \citenamefont
  {Keller}(2013)}]{friedan2013constraints}%
  \BibitemOpen
  \bibfield  {author} {\bibinfo {author} {\bibfnamefont {Daniel}\ \bibnamefont
  {Friedan}}\ and\ \bibinfo {author} {\bibfnamefont {Christoph~A}\ \bibnamefont
  {Keller}},\ }\bibfield  {title} {\enquote {\bibinfo {title} {Constraints on
  2d cft partition functions},}\ }\href@noop {} {\bibfield  {journal} {\bibinfo
   {journal} {Journal of High Energy Physics}\ }\textbf {\bibinfo {volume}
  {2013}},\ \bibinfo {pages} {180} (\bibinfo {year} {2013})}\BibitemShut
  {NoStop}%
\bibitem [{\citenamefont {Qualls}\ and\ \citenamefont
  {Shapere}(2014)}]{qualls2014bounds}%
  \BibitemOpen
  \bibfield  {author} {\bibinfo {author} {\bibfnamefont {Joshua~D}\
  \bibnamefont {Qualls}}\ and\ \bibinfo {author} {\bibfnamefont {Alfred~D}\
  \bibnamefont {Shapere}},\ }\bibfield  {title} {\enquote {\bibinfo {title}
  {Bounds on operator dimensions in 2d conformal field theories},}\ }\href@noop
  {} {\bibfield  {journal} {\bibinfo  {journal} {Journal of High Energy
  Physics}\ }\textbf {\bibinfo {volume} {2014}},\ \bibinfo {pages} {91}
  (\bibinfo {year} {2014})}\BibitemShut {NoStop}%
\bibitem [{\citenamefont {Hartman}\ \emph {et~al.}(2014)\citenamefont
  {Hartman}, \citenamefont {Keller},\ and\ \citenamefont
  {Stoica}}]{hartman2014universal}%
  \BibitemOpen
  \bibfield  {author} {\bibinfo {author} {\bibfnamefont {Thomas}\ \bibnamefont
  {Hartman}}, \bibinfo {author} {\bibfnamefont {Christoph~A}\ \bibnamefont
  {Keller}}, \ and\ \bibinfo {author} {\bibfnamefont {Bogdan}\ \bibnamefont
  {Stoica}},\ }\bibfield  {title} {\enquote {\bibinfo {title} {Universal
  spectrum of 2d conformal field theory in the large c limit},}\ }\href@noop {}
  {\bibfield  {journal} {\bibinfo  {journal} {Journal of High Energy Physics}\
  }\textbf {\bibinfo {volume} {2014}},\ \bibinfo {pages} {118} (\bibinfo {year}
  {2014})}\BibitemShut {NoStop}%
\bibitem [{\citenamefont {Qualls}(2015)}]{qualls2015universal}%
  \BibitemOpen
  \bibfield  {author} {\bibinfo {author} {\bibfnamefont {Joshua~D}\
  \bibnamefont {Qualls}},\ }\bibfield  {title} {\enquote {\bibinfo {title}
  {Universal bounds on operator dimensions in general 2d conformal field
  theories},}\ }\href@noop {} {\bibfield  {journal} {\bibinfo  {journal} {arXiv
  preprint arXiv:1508.00548}\ } (\bibinfo {year} {2015})}\BibitemShut {NoStop}%
\bibitem [{\citenamefont {Kim}\ \emph {et~al.}(2016)\citenamefont {Kim},
  \citenamefont {Kravchuk},\ and\ \citenamefont {Ooguri}}]{kim2016reflections}%
  \BibitemOpen
  \bibfield  {author} {\bibinfo {author} {\bibfnamefont {Hyungrok}\
  \bibnamefont {Kim}}, \bibinfo {author} {\bibfnamefont {Petr}\ \bibnamefont
  {Kravchuk}}, \ and\ \bibinfo {author} {\bibfnamefont {Hirosi}\ \bibnamefont
  {Ooguri}},\ }\bibfield  {title} {\enquote {\bibinfo {title} {Reflections on
  conformal spectra},}\ }\href@noop {} {\bibfield  {journal} {\bibinfo
  {journal} {Journal of High Energy Physics}\ }\textbf {\bibinfo {volume}
  {2016}},\ \bibinfo {pages} {184} (\bibinfo {year} {2016})}\BibitemShut
  {NoStop}%
\bibitem [{\citenamefont {Lin}\ \emph {et~al.}(2017)\citenamefont {Lin},
  \citenamefont {Shao}, \citenamefont {Wang},\ and\ \citenamefont
  {Yin}}]{lin20172}%
  \BibitemOpen
  \bibfield  {author} {\bibinfo {author} {\bibfnamefont {Ying-Hsuan}\
  \bibnamefont {Lin}}, \bibinfo {author} {\bibfnamefont {Shu-Heng}\
  \bibnamefont {Shao}}, \bibinfo {author} {\bibfnamefont {Yifan}\ \bibnamefont
  {Wang}}, \ and\ \bibinfo {author} {\bibfnamefont {Xi}~\bibnamefont {Yin}},\
  }\bibfield  {title} {\enquote {\bibinfo {title} {(2, 2) superconformal
  bootstrap in two dimensions},}\ }\href@noop {} {\bibfield  {journal}
  {\bibinfo  {journal} {Journal of High Energy Physics}\ }\textbf {\bibinfo
  {volume} {2017}},\ \bibinfo {pages} {112} (\bibinfo {year}
  {2017})}\BibitemShut {NoStop}%
\bibitem [{\citenamefont {Anous}\ \emph {et~al.}(2018)\citenamefont {Anous},
  \citenamefont {Mahajan},\ and\ \citenamefont
  {Shaghoulian}}]{anous2018parity}%
  \BibitemOpen
  \bibfield  {author} {\bibinfo {author} {\bibfnamefont {Tarek}\ \bibnamefont
  {Anous}}, \bibinfo {author} {\bibfnamefont {Raghu}\ \bibnamefont {Mahajan}},
  \ and\ \bibinfo {author} {\bibfnamefont {Edgar}\ \bibnamefont
  {Shaghoulian}},\ }\bibfield  {title} {\enquote {\bibinfo {title} {Parity and
  the modular bootstrap},}\ }\href@noop {} {\bibfield  {journal} {\bibinfo
  {journal} {SciPost Phys}\ }\textbf {\bibinfo {volume} {5}},\ \bibinfo {pages}
  {022} (\bibinfo {year} {2018})}\BibitemShut {NoStop}%
\bibitem [{\citenamefont {Collier}\ \emph {et~al.}(2018)\citenamefont
  {Collier}, \citenamefont {Lin},\ and\ \citenamefont
  {Yin}}]{collier2018modular}%
  \BibitemOpen
  \bibfield  {author} {\bibinfo {author} {\bibfnamefont {Scott}\ \bibnamefont
  {Collier}}, \bibinfo {author} {\bibfnamefont {Ying-Hsuan}\ \bibnamefont
  {Lin}}, \ and\ \bibinfo {author} {\bibfnamefont {Xi}~\bibnamefont {Yin}},\
  }\bibfield  {title} {\enquote {\bibinfo {title} {Modular bootstrap
  revisited},}\ }\href@noop {} {\bibfield  {journal} {\bibinfo  {journal}
  {Journal of High Energy Physics}\ }\textbf {\bibinfo {volume} {2018}},\
  \bibinfo {pages} {61} (\bibinfo {year} {2018})}\BibitemShut {NoStop}%
\bibitem [{\citenamefont {Afkhami-Jeddi}\ \emph {et~al.}(2019)\citenamefont
  {Afkhami-Jeddi}, \citenamefont {Hartman},\ and\ \citenamefont
  {Tajdini}}]{afkhami2019fast}%
  \BibitemOpen
  \bibfield  {author} {\bibinfo {author} {\bibfnamefont {Nima}\ \bibnamefont
  {Afkhami-Jeddi}}, \bibinfo {author} {\bibfnamefont {Thomas}\ \bibnamefont
  {Hartman}}, \ and\ \bibinfo {author} {\bibfnamefont {Amirhossein}\
  \bibnamefont {Tajdini}},\ }\bibfield  {title} {\enquote {\bibinfo {title}
  {Fast conformal bootstrap and constraints on 3d gravity},}\ }\href@noop {}
  {\bibfield  {journal} {\bibinfo  {journal} {Journal of High Energy Physics}\
  }\textbf {\bibinfo {volume} {2019}},\ \bibinfo {pages} {87} (\bibinfo {year}
  {2019})}\BibitemShut {NoStop}%
\bibitem [{\citenamefont {Cho}\ \emph {et~al.}(2019)\citenamefont {Cho},
  \citenamefont {Collier},\ and\ \citenamefont {Yin}}]{cho2019genus}%
  \BibitemOpen
  \bibfield  {author} {\bibinfo {author} {\bibfnamefont {Minjae}\ \bibnamefont
  {Cho}}, \bibinfo {author} {\bibfnamefont {Scott}\ \bibnamefont {Collier}}, \
  and\ \bibinfo {author} {\bibfnamefont {Xi}~\bibnamefont {Yin}},\ }\bibfield
  {title} {\enquote {\bibinfo {title} {Genus two modular bootstrap},}\
  }\href@noop {} {\bibfield  {journal} {\bibinfo  {journal} {Journal of High
  Energy Physics}\ }\textbf {\bibinfo {volume} {2019}},\ \bibinfo {pages} {22}
  (\bibinfo {year} {2019})}\BibitemShut {NoStop}%
\bibitem [{\citenamefont {Hartman}\ \emph {et~al.}(2019)\citenamefont
  {Hartman}, \citenamefont {Maz{\'a}{\v{c}}},\ and\ \citenamefont
  {Rastelli}}]{hartman2019sphere}%
  \BibitemOpen
  \bibfield  {author} {\bibinfo {author} {\bibfnamefont {Thomas}\ \bibnamefont
  {Hartman}}, \bibinfo {author} {\bibfnamefont {Dalimil}\ \bibnamefont
  {Maz{\'a}{\v{c}}}}, \ and\ \bibinfo {author} {\bibfnamefont {Leonardo}\
  \bibnamefont {Rastelli}},\ }\bibfield  {title} {\enquote {\bibinfo {title}
  {Sphere packing and quantum gravity},}\ }\href@noop {} {\bibfield  {journal}
  {\bibinfo  {journal} {Journal of High Energy Physics}\ }\textbf {\bibinfo
  {volume} {2019}},\ \bibinfo {pages} {48} (\bibinfo {year}
  {2019})}\BibitemShut {NoStop}%
\bibitem [{\citenamefont {Afkhami-Jeddi}\ \emph
  {et~al.}(2020{\natexlab{a}})\citenamefont {Afkhami-Jeddi}, \citenamefont
  {Cohn}, \citenamefont {Hartman}, \citenamefont {de~Laat},\ and\ \citenamefont
  {Tajdini}}]{afkhami2020high}%
  \BibitemOpen
  \bibfield  {author} {\bibinfo {author} {\bibfnamefont {Nima}\ \bibnamefont
  {Afkhami-Jeddi}}, \bibinfo {author} {\bibfnamefont {Henry}\ \bibnamefont
  {Cohn}}, \bibinfo {author} {\bibfnamefont {Thomas}\ \bibnamefont {Hartman}},
  \bibinfo {author} {\bibfnamefont {David}\ \bibnamefont {de~Laat}}, \ and\
  \bibinfo {author} {\bibfnamefont {Amirhossein}\ \bibnamefont {Tajdini}},\
  }\bibfield  {title} {\enquote {\bibinfo {title} {High-dimensional sphere
  packing and the modular bootstrap},}\ }\href@noop {} {\bibfield  {journal}
  {\bibinfo  {journal} {arXiv preprint arXiv:2006.02560}\ } (\bibinfo {year}
  {2020}{\natexlab{a}})}\BibitemShut {NoStop}%
\bibitem [{\citenamefont {Afkhami-Jeddi}\ \emph
  {et~al.}(2020{\natexlab{b}})\citenamefont {Afkhami-Jeddi}, \citenamefont
  {Cohn}, \citenamefont {Hartman},\ and\ \citenamefont
  {Tajdini}}]{afkhami2020free}%
  \BibitemOpen
  \bibfield  {author} {\bibinfo {author} {\bibfnamefont {Nima}\ \bibnamefont
  {Afkhami-Jeddi}}, \bibinfo {author} {\bibfnamefont {Henry}\ \bibnamefont
  {Cohn}}, \bibinfo {author} {\bibfnamefont {Thomas}\ \bibnamefont {Hartman}},
  \ and\ \bibinfo {author} {\bibfnamefont {Amirhossein}\ \bibnamefont
  {Tajdini}},\ }\bibfield  {title} {\enquote {\bibinfo {title} {Free partition
  functions and an averaged holographic duality},}\ }\href@noop {} {\bibfield
  {journal} {\bibinfo  {journal} {arXiv preprint arXiv:2006.04839}\ } (\bibinfo
  {year} {2020}{\natexlab{b}})}\BibitemShut {NoStop}%
\bibitem [{\citenamefont {Gliozzi}(2020)}]{gliozzi2020modular}%
  \BibitemOpen
  \bibfield  {author} {\bibinfo {author} {\bibfnamefont {Ferdinando}\
  \bibnamefont {Gliozzi}},\ }\bibfield  {title} {\enquote {\bibinfo {title}
  {Modular bootstrap, elliptic points, and quantum gravity},}\ }\href@noop {}
  {\bibfield  {journal} {\bibinfo  {journal} {Physical Review Research}\
  }\textbf {\bibinfo {volume} {2}},\ \bibinfo {pages} {013327} (\bibinfo {year}
  {2020})}\BibitemShut {NoStop}%
\bibitem [{\citenamefont {Schellekens}(1993)}]{schellekens1992meromorphic}%
  \BibitemOpen
  \bibfield  {author} {\bibinfo {author} {\bibfnamefont {A.~N.}\ \bibnamefont
  {Schellekens}},\ }\bibfield  {title} {\enquote {\bibinfo {title} {Meromorphic
  c = 24 conformal field theories,},}\ }\href@noop {} {\bibfield  {journal}
  {\bibinfo  {journal} {Commun. Math. Phys.}\ }\textbf {\bibinfo {volume}
  {153}},\ \bibinfo {pages} {159} (\bibinfo {year} {1993})}\BibitemShut
  {NoStop}%
\bibitem [{\citenamefont {Mukhi}(2019)}]{mukhi2019classification}%
  \BibitemOpen
  \bibfield  {author} {\bibinfo {author} {\bibfnamefont {Sunil}\ \bibnamefont
  {Mukhi}},\ }\bibfield  {title} {\enquote {\bibinfo {title} {Classification of
  rcft from holomorphic modular bootstrap: A status report},}\ }\href@noop {}
  {\bibfield  {journal} {\bibinfo  {journal} {arXiv preprint arXiv:1910.02973}\
  } (\bibinfo {year} {2019})},\ \bibinfo {note} {and references
  therein}\BibitemShut {NoStop}%
\bibitem [{\citenamefont {Schlingemann}\ and\ \citenamefont
  {Werner}(2001)}]{schlingemann2001quantum}%
  \BibitemOpen
  \bibfield  {author} {\bibinfo {author} {\bibfnamefont {Dirk}\ \bibnamefont
  {Schlingemann}}\ and\ \bibinfo {author} {\bibfnamefont {Reinhard~F}\
  \bibnamefont {Werner}},\ }\bibfield  {title} {\enquote {\bibinfo {title}
  {Quantum error-correcting codes associated with graphs},}\ }\href@noop {}
  {\bibfield  {journal} {\bibinfo  {journal} {Physical Review A}\ }\textbf
  {\bibinfo {volume} {65}},\ \bibinfo {pages} {012308} (\bibinfo {year}
  {2001})}\BibitemShut {NoStop}%
\bibitem [{\citenamefont {Schlingemann}(2002)}]{schlingemann2002stabilizer}%
  \BibitemOpen
  \bibfield  {author} {\bibinfo {author} {\bibfnamefont {D}~\bibnamefont
  {Schlingemann}},\ }\bibfield  {title} {\enquote {\bibinfo {title} {Stabilizer
  codes can be realized as graph codes},}\ }\href@noop {} {\bibfield  {journal}
  {\bibinfo  {journal} {Quantum Information \& Computation}\ }\textbf {\bibinfo
  {volume} {2}},\ \bibinfo {pages} {307--323} (\bibinfo {year}
  {2002})}\BibitemShut {NoStop}%
\bibitem [{\citenamefont {D{\"u}r}\ \emph {et~al.}(2003)\citenamefont
  {D{\"u}r}, \citenamefont {Aschauer},\ and\ \citenamefont
  {Briegel}}]{dur2003multiparticle}%
  \BibitemOpen
  \bibfield  {author} {\bibinfo {author} {\bibfnamefont {Wolfgang}\
  \bibnamefont {D{\"u}r}}, \bibinfo {author} {\bibfnamefont {Hans}\
  \bibnamefont {Aschauer}}, \ and\ \bibinfo {author} {\bibfnamefont {H-J}\
  \bibnamefont {Briegel}},\ }\bibfield  {title} {\enquote {\bibinfo {title}
  {Multiparticle entanglement purification for graph states},}\ }\href@noop {}
  {\bibfield  {journal} {\bibinfo  {journal} {Physical review letters}\
  }\textbf {\bibinfo {volume} {91}},\ \bibinfo {pages} {107903} (\bibinfo
  {year} {2003})}\BibitemShut {NoStop}%
\bibitem [{\citenamefont {Glynn}\ \emph {et~al.}(2004)\citenamefont {Glynn},
  \citenamefont {Gulliver}, \citenamefont {Maks},\ and\ \citenamefont
  {Gupta}}]{glynn2004geometry}%
  \BibitemOpen
  \bibfield  {author} {\bibinfo {author} {\bibfnamefont {David~G}\ \bibnamefont
  {Glynn}}, \bibinfo {author} {\bibfnamefont {T~Aaron}\ \bibnamefont
  {Gulliver}}, \bibinfo {author} {\bibfnamefont {Johannes~G}\ \bibnamefont
  {Maks}}, \ and\ \bibinfo {author} {\bibfnamefont {Manish~K}\ \bibnamefont
  {Gupta}},\ }\bibfield  {title} {\enquote {\bibinfo {title} {The geometry of
  additive quantum codes},}\ }\href
  {https://www.researchgate.net/publication/230899738_The_geometry_of_additive_quantum_codes.}
  {\  (\bibinfo {year} {2004})}\BibitemShut {NoStop}%
\bibitem [{\citenamefont {Van~den Nest}\ \emph {et~al.}(2004)\citenamefont
  {Van~den Nest}, \citenamefont {Dehaene},\ and\ \citenamefont
  {De~Moor}}]{van2004graphical}%
  \BibitemOpen
  \bibfield  {author} {\bibinfo {author} {\bibfnamefont {Maarten}\ \bibnamefont
  {Van~den Nest}}, \bibinfo {author} {\bibfnamefont {Jeroen}\ \bibnamefont
  {Dehaene}}, \ and\ \bibinfo {author} {\bibfnamefont {Bart}\ \bibnamefont
  {De~Moor}},\ }\bibfield  {title} {\enquote {\bibinfo {title} {Graphical
  description of the action of local clifford transformations on graph
  states},}\ }\href@noop {} {\bibfield  {journal} {\bibinfo  {journal}
  {Physical Review A}\ }\textbf {\bibinfo {volume} {69}},\ \bibinfo {pages}
  {022316} (\bibinfo {year} {2004})}\BibitemShut {NoStop}%
\bibitem [{\citenamefont {Hein}\ \emph {et~al.}(2006)\citenamefont {Hein},
  \citenamefont {D{\"u}r}, \citenamefont {Eisert}, \citenamefont {Raussendorf},
  \citenamefont {Nest},\ and\ \citenamefont {Briegel}}]{hein2006entanglement}%
  \BibitemOpen
  \bibfield  {author} {\bibinfo {author} {\bibfnamefont {Marc}\ \bibnamefont
  {Hein}}, \bibinfo {author} {\bibfnamefont {Wolfgang}\ \bibnamefont
  {D{\"u}r}}, \bibinfo {author} {\bibfnamefont {Jens}\ \bibnamefont {Eisert}},
  \bibinfo {author} {\bibfnamefont {Robert}\ \bibnamefont {Raussendorf}},
  \bibinfo {author} {\bibfnamefont {M}~\bibnamefont {Nest}}, \ and\ \bibinfo
  {author} {\bibfnamefont {H-J}\ \bibnamefont {Briegel}},\ }\bibfield  {title}
  {\enquote {\bibinfo {title} {Entanglement in graph states and its
  applications},}\ }\href@noop {} {\bibfield  {journal} {\bibinfo  {journal}
  {arXiv preprint quant-ph/0602096}\ } (\bibinfo {year} {2006})}\BibitemShut
  {NoStop}%
\bibitem [{\citenamefont {MacWilliams}\ \emph {et~al.}(1978)\citenamefont
  {MacWilliams}, \citenamefont {Odlyzko}, \citenamefont {Sloane},\ and\
  \citenamefont {Ward}}]{macwilliams1978self}%
  \BibitemOpen
  \bibfield  {author} {\bibinfo {author} {\bibfnamefont {F~Jessie}\
  \bibnamefont {MacWilliams}}, \bibinfo {author} {\bibfnamefont {Andrew~M.}\
  \bibnamefont {Odlyzko}}, \bibinfo {author} {\bibfnamefont {Neil~JA}\
  \bibnamefont {Sloane}}, \ and\ \bibinfo {author} {\bibfnamefont {Harold~N.}\
  \bibnamefont {Ward}},\ }\bibfield  {title} {\enquote {\bibinfo {title}
  {Self-dual codes over gf (4)},}\ }\href@noop {} {\bibfield  {journal}
  {\bibinfo  {journal} {Journal of Combinatorial Theory, Series A}\ }\textbf
  {\bibinfo {volume} {25}},\ \bibinfo {pages} {288--318} (\bibinfo {year}
  {1978})}\BibitemShut {NoStop}%
\bibitem [{\citenamefont {Shor}\ and\ \citenamefont
  {Laflamme}(1997)}]{shor1997quantum}%
  \BibitemOpen
  \bibfield  {author} {\bibinfo {author} {\bibfnamefont {Peter}\ \bibnamefont
  {Shor}}\ and\ \bibinfo {author} {\bibfnamefont {Raymond}\ \bibnamefont
  {Laflamme}},\ }\bibfield  {title} {\enquote {\bibinfo {title} {Quantum analog
  of the macwilliams identities for classical coding theory},}\ }\href@noop {}
  {\bibfield  {journal} {\bibinfo  {journal} {Physical review letters}\
  }\textbf {\bibinfo {volume} {78}},\ \bibinfo {pages} {1600} (\bibinfo {year}
  {1997})}\BibitemShut {NoStop}%
\bibitem [{\citenamefont {Nebe}\ \emph {et~al.}(2006)\citenamefont {Nebe},
  \citenamefont {Rains},\ and\ \citenamefont {Sloane}}]{nebe2006self}%
  \BibitemOpen
  \bibfield  {author} {\bibinfo {author} {\bibfnamefont {Gabriele}\
  \bibnamefont {Nebe}}, \bibinfo {author} {\bibfnamefont {Eric~M}\ \bibnamefont
  {Rains}}, \ and\ \bibinfo {author} {\bibfnamefont {Neil James~Alexander}\
  \bibnamefont {Sloane}},\ }\href@noop {} {\emph {\bibinfo {title} {Self-dual
  codes and invariant theory}}},\ Vol.~\bibinfo {volume} {17}\ (\bibinfo
  {publisher} {Springer},\ \bibinfo {year} {2006})\BibitemShut {NoStop}%
\bibitem [{\citenamefont {Van~den Nest}\ and\ \citenamefont
  {De~Moor}(2005)}]{van2005edge}%
  \BibitemOpen
  \bibfield  {author} {\bibinfo {author} {\bibfnamefont {Maarten}\ \bibnamefont
  {Van~den Nest}}\ and\ \bibinfo {author} {\bibfnamefont {Bart}\ \bibnamefont
  {De~Moor}},\ }\bibfield  {title} {\enquote {\bibinfo {title} {Edge-local
  equivalence of graphs},}\ }\href@noop {} {\bibfield  {journal} {\bibinfo
  {journal} {arXiv preprint math/0510246}\ } (\bibinfo {year}
  {2005})}\BibitemShut {NoStop}%
\bibitem [{\citenamefont {Danielsen}\ and\ \citenamefont
  {Parker}(2006)}]{danielsen2006classification}%
  \BibitemOpen
  \bibfield  {author} {\bibinfo {author} {\bibfnamefont {Lars~Eirik}\
  \bibnamefont {Danielsen}}\ and\ \bibinfo {author} {\bibfnamefont {Matthew~G}\
  \bibnamefont {Parker}},\ }\bibfield  {title} {\enquote {\bibinfo {title} {On
  the classification of all self-dual additive codes over gf (4) of length up
  to 12},}\ }\href@noop {} {\bibfield  {journal} {\bibinfo  {journal} {Journal
  of Combinatorial Theory, Series A}\ }\textbf {\bibinfo {volume} {113}},\
  \bibinfo {pages} {1351--1367} (\bibinfo {year} {2006})}\BibitemShut {NoStop}%
\bibitem [{\citenamefont {Dymarsky}\ and\ \citenamefont {Shapere}(2020)}]{QLC}%
  \BibitemOpen
  \bibfield  {author} {\bibinfo {author} {\bibfnamefont {Anatoly}\ \bibnamefont
  {Dymarsky}}\ and\ \bibinfo {author} {\bibfnamefont {Alfred}\ \bibnamefont
  {Shapere}},\ }\bibfield  {title} {\enquote {\bibinfo {title} {Quantum
  stabilizer codes, lattices, and cfts},}\ }\href@noop {} {\bibfield  {journal}
  {\bibinfo  {journal} {arXiv preprint arXiv:2009.01244}\ } (\bibinfo {year}
  {2020})}\BibitemShut {NoStop}%
\bibitem [{\citenamefont {Conway}(1997)}]{conway1997sensual}%
  \BibitemOpen
  \bibfield  {author} {\bibinfo {author} {\bibfnamefont {John~Horton}\
  \bibnamefont {Conway}},\ }\href@noop {} {\emph {\bibinfo {title} {The sensual
  (quadratic) form}}},\ Vol.~\bibinfo {volume} {26}\ (\bibinfo  {publisher}
  {American Mathematical Soc.},\ \bibinfo {year} {1997})\BibitemShut {NoStop}%
\bibitem [{\citenamefont {Kraus}\ and\ \citenamefont
  {Larsen}(2007)}]{kraus2007partition}%
  \BibitemOpen
  \bibfield  {author} {\bibinfo {author} {\bibfnamefont {Per}\ \bibnamefont
  {Kraus}}\ and\ \bibinfo {author} {\bibfnamefont {Finn}\ \bibnamefont
  {Larsen}},\ }\bibfield  {title} {\enquote {\bibinfo {title} {Partition
  functions and elliptic genera from supergravity},}\ }\href@noop {} {\bibfield
   {journal} {\bibinfo  {journal} {Journal of High Energy Physics}\ }\textbf
  {\bibinfo {volume} {2007}},\ \bibinfo {pages} {002} (\bibinfo {year}
  {2007})}\BibitemShut {NoStop}%
\bibitem [{fut()}]{future}%
  \BibitemOpen
  \href@noop {} {}\bibinfo {note} {Work in progress}\BibitemShut {NoStop}%
\bibitem [{\citenamefont {Witten}(1989)}]{witten1989quantum}%
  \BibitemOpen
  \bibfield  {author} {\bibinfo {author} {\bibfnamefont {Edward}\ \bibnamefont
  {Witten}},\ }\bibfield  {title} {\enquote {\bibinfo {title} {Quantum field
  theory and the jones polynomial},}\ }\href@noop {} {\bibfield  {journal}
  {\bibinfo  {journal} {Communications in Mathematical Physics}\ }\textbf
  {\bibinfo {volume} {121}},\ \bibinfo {pages} {351--399} (\bibinfo {year}
  {1989})}\BibitemShut {NoStop}%
\bibitem [{\citenamefont {Ginsparg}(1987)}]{PhysRevD.35.648}%
  \BibitemOpen
  \bibfield  {author} {\bibinfo {author} {\bibfnamefont {Paul}\ \bibnamefont
  {Ginsparg}},\ }\bibfield  {title} {\enquote {\bibinfo {title} {On toroidal
  compactification of heterotic superstrings},}\ }\href {\doibase
  10.1103/PhysRevD.35.648} {\bibfield  {journal} {\bibinfo  {journal} {Phys.
  Rev. D}\ }\textbf {\bibinfo {volume} {35}},\ \bibinfo {pages} {648--654}
  (\bibinfo {year} {1987})}\BibitemShut {NoStop}%
\bibitem [{\citenamefont {Narain}(1986)}]{NARAIN198641}%
  \BibitemOpen
  \bibfield  {author} {\bibinfo {author} {\bibfnamefont {K.S.}\ \bibnamefont
  {Narain}},\ }\bibfield  {title} {\enquote {\bibinfo {title} {New heterotic
  string theories in uncompactified dimensions $<$ 10},}\ }\href {\doibase
  https://doi.org/10.1016/0370-2693(86)90682-9} {\bibfield  {journal} {\bibinfo
   {journal} {Physics Letters B}\ }\textbf {\bibinfo {volume} {169}},\ \bibinfo
  {pages} {41 -- 46} (\bibinfo {year} {1986})}\BibitemShut {NoStop}%
\bibitem [{\citenamefont {Narain}\ \emph {et~al.}(1987)\citenamefont {Narain},
  \citenamefont {Sarmadi},\ and\ \citenamefont {Witten}}]{NARAIN1987369}%
  \BibitemOpen
  \bibfield  {author} {\bibinfo {author} {\bibfnamefont {K.S.}\ \bibnamefont
  {Narain}}, \bibinfo {author} {\bibfnamefont {M.H.}\ \bibnamefont {Sarmadi}},
  \ and\ \bibinfo {author} {\bibfnamefont {E.}~\bibnamefont {Witten}},\
  }\bibfield  {title} {\enquote {\bibinfo {title} {A note on toroidal
  compactification of heterotic string theory},}\ }\href {\doibase
  https://doi.org/10.1016/0550-3213(87)90001-0} {\bibfield  {journal} {\bibinfo
   {journal} {Nuclear Physics B}\ }\textbf {\bibinfo {volume} {279}},\ \bibinfo
  {pages} {369 -- 379} (\bibinfo {year} {1987})}\BibitemShut {NoStop}%
\bibitem [{\citenamefont {Witten}(2007)}]{witten2007three}%
  \BibitemOpen
  \bibfield  {author} {\bibinfo {author} {\bibfnamefont {Edward}\ \bibnamefont
  {Witten}},\ }\bibfield  {title} {\enquote {\bibinfo {title}
  {Three-dimensional gravity revisited},}\ }\href@noop {} {\bibfield  {journal}
  {\bibinfo  {journal} {arXiv preprint arXiv:0706.3359}\ } (\bibinfo {year}
  {2007})}\BibitemShut {NoStop}%
\end{thebibliography}%

%\end{document}
\newpage

\section{Supplemental Material}
\section{Grand canonical partition function}
For 2d CFTs with  $U(1)^n\times U(1)^n$ symmetry  compactified on a torus is it natural to introduce the grand canonical partition function, which besides the modular parameter $\tau$, would also depend on $2n$ chemical potentials
\bea
Z(\tau,\bar \tau,\vec{\xi},\vec{\bar \xi}) =\Tr\left(q^{L_0-{c\over 24}}{\bar q}^{{\bar L}_0-{c\over 24}}e^{2\pi i \vec{\xi}\cdot \vec{q} +2\pi i \vec{\bar \xi}\cdot \vec{\bar q} }\right),
\eea
where $q=e^{i\pi \tau}, \bar q=e^{-i\pi \bar \tau}$ and $\vec{Q},\vec{\bar Q}$ are the  $U(1)$ charges.
For the Narain CFT associated with the lattice $\Lambda$ it can be represented as follows 
\bea
\label{GC}
\Theta_\Lambda&=&\sum_{(p_L,p_R)\in \Lambda} q^{p_L^2/2} {\bar q}^{p_R^2/2} e^{2\pi i \vec{p}_L \cdot \vec{\xi}} e^{2\pi i \vec{p}_R \cdot \vec{\bar\xi}},\\ 
Z&=&{\Theta_\Lambda\over |\eta(\tau)\eta(\bar \tau)|^n}. 
\eea
Under modular transformations $Z$ changes covariantly
\bea
Z(\tau+1,\bar \tau+1,\vec{\xi}, \vec{\bar \xi})&=&Z(\tau,\bar \tau,\vec{\xi}, \vec{\bar \xi}),\\
Z(-{1\over \tau},-{1\over \bar \tau},{{\vec \xi}\over \tau},{\vec{\bar \xi}\over \bar \tau})&=&Z(\tau,\bar \tau, \vec{\xi}, \vec{\bar \xi}) e^{i\pi ({\vec{\xi}^2\over \tau}-{\vec{\bar\xi}^2\over \bar \tau})}. \label{mtrans}
\eea
$Z$ is related to the worldsheet path integral $Z_{PI}$ as follows:
\bea
Z=Z_{\rm PI}\, e^{-{\pi (\vec{\xi}+\vec{\bar \xi})^2\over 2\tau_2}}. \label{ZZPI}
\eea
The path integral is invariant under modular transformations. The relation between $Z$ and $Z_{PI}$ is discussed in detail in the Appendix A of \cite{kraus2007partition}. 

The full group of T-dualities includes orthogonal transformations $O_L \times O_R \in O(n) \times O(n)$, which change the lattice $\Lambda$. While the conventional partition function is invariant under $O(n) \times O(n)$,  the grand canonical partition function changes by a rotation of $\vec{\xi},\vec{\bar \xi}$. Hence, we should recognize $Z(\tau,\bar \tau,\vec{\xi},\vec{\bar \xi})$ and $Z'(\tau,\bar \tau,\vec{\xi},\vec{\bar \xi})$ as physically equivalent if 
\bea
Z(\tau,\bar \tau,\vec{\xi},\vec{\bar \xi})=Z'(\tau,\bar \tau,O_L \vec{\xi},O_R \vec{\bar \xi})
\label{tdualityZ}
\eea
for some orthogonal $O_L,O_R$.

\section{$Z(\tau,\bar \tau,\vec{\xi},\vec{\bar \xi})$ for code theories}
For code theories the grand canonical partition function \eqref{GC} drastically simplifies when all components of $\vec{\xi}$ and $\vec{\bar \xi}$ are equal
\bea
\xi^I=\xi,\qquad \bar \xi^I=\bar \xi.
\eea
This choice restores permutation symmetry between different $U(1)$ generators.  In the language of codes this is the symmetry of permutations of qubits. In this case $Z_\C$ -- the grand canonical partition function of a code CFT associated with the code $\C$ -- will only depend on such aggregate properties of codewords as the total number of different letters, i.e.~precisely the information contained in the full enumerator polynomial 
\bea
\label{FEP}
W_\C(t,x,y,z)=\sum_{c \in \C } t^{n-w(c)} x^{ w_x(c)} y^{ w_y(c)} z^{ w_z(c)},\\ w(c)=w_x(c)+w_y(c)+w_z(c).
\eea
Here $w_x(c)$ is the number of generators $\sigma_x$ contained in the stabilizer element (codeword) $c$, etc.  The refined enumerator polynomial, eq.~(5) in the main text, is related to the full enumerator polynomial \eqref{FEP} as follows
\bea
W_\C(x,y,z)=W_\C(x,z,y,z).
\eea
An explicit calculation shows that $\Theta$ is then  given by 
\bea
\nonumber
\Theta(\tau,\bar\tau,\xi,\bar \xi)={W_\C\left(b \bar b+c\bar c, a\bar a+d\bar d, b \bar b-c\bar c, a\bar a-d\bar d\right)}, \\
\label{GCT}
\eea
giving rise to eq.~(21) in the main text. Here 
\bea
\nonumber
a&=&\theta_2(\tau,\xi),\, \quad b=\theta_3(\tau,\xi),\,\quad c=\theta_4(\tau,\xi),\, \quad d=\theta_1(\tau,\xi),\\
\bar a&=&\theta_2(-\bar \tau,\bar \xi),\,\,  \bar b=\theta_3(-\bar \tau,\bar \xi),\,\, \bar c=\theta_4(-\bar \tau,\bar \xi),\,\, \bar d=\theta_1(-\bar \tau,\bar \xi),\nonumber
\eea
and 
\bea
\theta_1(\tau,z)&=&-i\sum_{n \in\Z} e^{i \pi \tau (n+1/2)^2+2\pi i (n+1/2) z}(-1)^n,\\
\theta_2(\tau,z)&=&\sum_{n \in\Z} e^{i \pi \tau (n+1/2)^2+2\pi i (n+1/2) z},\\
\theta_3(\tau,z)&=&\sum_{n \in\Z} e^{i \pi \tau n^2+2\pi i n z},\\
\theta_4(\tau,z)&=&\sum_{n \in\Z} e^{i \pi \tau n^2+2\pi i n z} (-1)^n.
\eea
It is illustrative to discuss the modular covariance of $Z$ for code theories. Under the transformation $\tau\rightarrow \tau+1$, $\bar \tau\rightarrow \bar\tau+1$, the variables
$a,d$ acquire a phase $e^{i\pi/4}$ and $\bar a,\bar d$ acquire the opposite phase $e^{-i\pi/4}$. This transformation also exchanges $b \leftrightarrow  c$ and $\bar b \leftrightarrow  \bar c$. 
Therefore  \eqref{GCT} remains invariant, as follows from the fact that the code $\C$ is real, $W_\C(t,x,y,z)=W_\C(t,x,-y,z)$, see \cite{QLC} for details.  Under the transformation $\tau \rightarrow -1/\tau$, $\bar \tau \rightarrow -1/\bar \tau$, $\xi \rightarrow \xi/\tau$, $\bar \xi \rightarrow \bar \xi/\bar \tau$ the change is as follows
\bea
a& \rightarrow&\sqrt{-i \tau}e^{i\pi  \xi^2\over \tau} c,\\
b& \rightarrow&\sqrt{-i \tau}e^{i\pi  \xi^2\over \tau} b,\\
c &\rightarrow&\sqrt{-i \tau}e^{i\pi  \xi^2\over \tau} a,\\
d &\rightarrow&\sqrt{-i \tau}e^{i\pi  \xi^2\over \tau} (-id),
\eea
and 
\bea
\bar a& \rightarrow&\sqrt{i \bar\tau}e^{-{i\pi  \bar \xi^2\over \bar\tau}} \bar c,\\
\bar b& \rightarrow&\sqrt{i \bar\tau}e^{-{i\pi  \bar \xi^2\over \bar\tau}} \bar b,\\
\bar c &\rightarrow&\sqrt{i \bar \tau}e^{-{i\pi  \bar \xi^2\over \bar\tau}} \bar a,\\
\bar d &\rightarrow&\sqrt{i \bar \tau}e^{-{i\pi  \bar \xi^2\over \bar\tau}} (i \bar d).
\eea
Therefore $\Theta$ changes covariantly, as outlined in \eqref{mtrans},
\bea
\Theta \rightarrow (\tau \bar \tau)^{n/2} e^{in\pi \xi^2\over \tau} e^{-{i n\pi \bar\xi^2\over \bar \tau}}  \Theta ,
\eea
which follows from the invariance of $W_\C(t,x,y,z)$ under MacWilliams identity \cite{nebe2006self}
\bea
t\rightarrow {t+x+y+z\over 2},\quad
x\rightarrow {t+x-y-z\over 2},\\
y\rightarrow {t-x+y-z\over 2},\quad 
z\rightarrow {t-x-y+z\over 2}.
\eea
In other words, covariance of the grand canonical partition function under modular transformation follows from the properties of the full enumerator $W_\C(t,x,y,z)$ exactly in the same way that invariance of the partition function follows from the properties of the refined enumerator $W_\C(x,y,z)$.

To summarize, the full enumerator polynomial has a clear physical meaning in the context of the relation between codes and CFTs: it governs  the CFT grand canonical ensemble decorated by $U(1)$ charges with the values of chemical potentials invariant under qubit permutation symmetry. A natural question to ask is if the full enumerator polynomial can resolve or modify the presence of fake refined enumerator polynomials, i.e. modular invariant and otherwise appropriate $Z$ without known CFT counterparts. Unfortunately, using the full enumerator instead of the refined one does not change the picture:  there is still a rapidly growing number of fake full enumerator polynomials  not associated with any code starting from $n=3$. 

\section{$Z(\tau,\bar \tau,{\xi},{\bar \xi})$ for isospectral theories}
As summarized in \eqref{tdualityZ}, the grand canonical partition function changes covariantly under T-duality transformations.  The only T-dualities which map a code CFT into another code CFT are those whose action  reduces  to an action on codes, see Appendix D of \cite{QLC}. They permute qubits and exchange $\sigma_x$ and $\sigma_z$ in the codewords (stabilizer generators). Acting on $\xi^I, \bar \xi^I$ they may permute the components and flip the signs of some $\bar \xi^I \rightarrow -\bar \xi^I$. (Since $\theta_2,\theta_3,\theta_4$ are even and $\theta_1$ is odd with respect to the second argument, this flip is equivalent to  exchanging $\sigma_x$ and $\sigma_z$ at the level of codewords.) We shall regard pairs of $Z(\tau,\bar \tau,\vec{\xi},\vec{\bar \xi})$ which can be mapped to each other by a T-duality  transformation as equivalent. The question we would like to understand is whether, by using the grand canonical partition function with $\xi^I=\xi,\, \bar \xi^I=\bar \xi$ (or more accurately with $|\xi^I|=\xi,\, |\bar \xi^I|=\bar \xi$) to account for possible T-dualities, we can remove the degeneracy between isospectral theories. 

By using T-duality transformation any code (associated with a code CFT) discussed in this paper, and more generally in \cite{QLC}, can be brought to graph form, see eqs.~(2) and  (9) in the main text. This representation is not unique: there could be additional T-dualities, which map a graph code CFT to another graph code CFT. As we have discussed in the main text, at the level of graphs they act by Edge Local Complementation (ELC). These transformations leave the refined enumerator polynomial invariant but change the full enumerator polynomial. The corresponding $Z(\tau,\bar \tau, \xi, \bar \xi)$ would be different, but in fact would be related by a permutation of $\xi^I, \bar\xi^I$ and a sign flip of some $\bar \xi^I \rightarrow -\bar \xi^I$, although  this relation might not be obvious at the level of $Z(\tau,\bar \tau, \xi, \bar \xi)$. Importantly, these generate all transformations which are equivalences of $Z(\tau,\bar \tau, \xi, \bar \xi)$. Hence, to verify if a particular pair of  isospectral theories would have the same (equivalent) $Z(\tau,\bar \tau, \xi, \bar \xi)$, we must do the following. For each theory in question we should consider the full set of graphs associated with this theory and related by ELC to each other. Each graph in the ELC class defines a graph code and we need to evaluate its full enumerator polynomial. Of course, the full enumerator polynomial is invariant under qubit permutations (node permutations at the level of graphs), so only non-isomorphic graphs from the ELC class should be considered. Finally, one should compare full enumerator polynomials obtained in this way for both theories. If there is even one polynomial which appears for both theories, then the theories are isospectral not only in the sense of the partition function but also in the sense  of the grand canonical partition function. If the sets of polynomials associated with the two theories have no common elements, the grand canonical ensemble removes the degeneracy. 

We illustrate this general scheme for the pair of isospectral theories with $n=7$ discussed in the main text. Up to isomorphisms, the ELC equivalence class  of the first theory includes 9 graphs; there are 10 graphs for the second theory. All 19 full enumerator polynomials associated with these graphs turn out to be distinct. It is less important that all 9 (10) graphs of each ELC class lead to a unique full enumerator. It is important though that the resulting sets of  polynomials do not overlap. Hence in this particular case the grand canonical ensemble removes the degeneracy between these two theories. We have performed this calculation using the full classification of the ELC classes for $n=7$ graphs, see Appendix E of \cite{QLC}. Extending this analysis to instances of isospectral  theories with $n=8$ would require first classifying ELC orbits for graphs with $n=8$ nodes.

\end{document}